\documentstyle[12pt]{article}
\newcommand{\beq}{\begin{equation}}
\newcommand{\eeq}{\end{equation}}
\newcommand{\bea}{\begin{eqnarray}}
\newcommand{\eea}{\end{eqnarray}}

\newcommand{\k}{\kappa}

\newcommand{\rg}{\sqrt{g}}

\newcommand{\p}{\phi}

\newcommand{\tN}{\tilde{N}}
\renewcommand{\d}{\delta}
\renewcommand{\l}{\lambda}

\renewcommand{\b}{\beta}
\renewcommand{\a}{\alpha}
\newcommand{\E}{{\cal E}}
\newcommand{\ER}{\sqrt{\cal E}}

\newcommand{\A}{\mbox{\AE}}
\newcommand{\G}{{\cal G}}
\renewcommand{\H}{{\cal H}}
\newcommand{\N}{{\cal N}}

\newcommand{\n}{\nu}
\newcommand{\m}{\mu}
\newcommand{\r}{\rho}
\newcommand{\s}{\sigma}

\newcommand{\oh}{\frac{1}{2}}
\newcommand{\oq}{\frac{1}{4}}

\newcommand{\non}{\nonumber}

\renewcommand{\t}{\tau}
\newcommand{\rf}[1]{(\ref{#1})}
\newcommand{\ra}{\rightarrow}

\newcommand{\pa}{\partial}

\newcommand{\K}{\left[ -\int d^3x' \; {1\over 4\N \k^2} G_{ab}
                          \pa_\t q^a \pa_\t q^b \right]^{1/2} }
\newcommand{\V}{\left[ \int d^3x'' \; \N \rg U \right]^{1/2} }
\begin{document}

\addtolength{\baselineskip}{0.20\baselineskip}

\hfill NBI-HE-95-24

\hfill gr-qc/9508033

\hfill August 1995

\begin{center}

\vspace{24pt}

{ {\LARGE \bf Field Theory As Free Fall} }

\end{center}

\vspace{32pt}

\begin{center}

{\sl J. Greensite}\footnote{Permanent address: Physics and Astronomy
Dept., San Francisco State University, San Francisco CA 94132 USA}
\footnote{email: greensit@stars.sfsu.edu}

\vspace{12pt}

The Niels Bohr Institute \\
Blegdamsvej 17 \\
DK-2100 Copenhagen \O \\
Denmark \\

\vspace{96pt}

{\bf Abstract}

\end{center}

\bigskip

   It is shown that the classical field equations pertaining to
gravity coupled to other bosonic fields are equivalent to a
single geodesic equation, describing the free fall of a point particle
in superspace. Some implications for quantum gravity are discussed.

\vfill

\newpage

\section{Introduction}

   In the canonical formulation of general relativity, it is customary
to view the time-evolution of the three metric $g_{ij}(x)$, as well as the
evolution of any other non-gravitational fields $\p^A(x)$, as tracing
out a trajectory in the "space of all fields" known as superspace.
In this article I will show that the classical equations of motion
and constraints, which govern the dynamics of $g_{ij}$ and $\p^A$,
are equivalent to a geodesic equation in superspace, with a certain
supermetric that will be specified.  Field theory
in general relativity may therefore be regarded as describing the free
fall of a point particle in a (super) gravitational field.  Connections to
Jacobi's principle in mechanics, implications for quantization
(in particular, the problem of time in quantum gravity), and
possible generalizations, will also be discussed below.

\section{Geodesics in Superspace}

   Let $\{q^a(x),p_a(x),~a=1,2,...,n_f\}$ denote the canonical
variables of a set of
integer-spin fields including gravity, i.e. $\{q^a(x)\}=\{g_{ij}(x),
\phi^A(x)\}$, with the non-gravitational fields scaled by an appropriate
power of Newton's constant so as to be dimensionless.\footnote{I will work
here entirely in the metric ($g_{\m \n}$) formalism; spinor fields will
not be discussed.} The first-order ADM action has the form
\bea
      S &=& \int d^4x \; [p_a \pa_t q^a - N\H_x - N_i\H^i_x]
\non \\
      \H_x &=& \k^2 G^{ab} p_a p_b + \rg U(q)
\non \\
      \H^i_x &=& O^{ia}[q,\pa_x] p_a
\label{ADM}
\eea
leading to the dynamical equations and constraints
\bea
  \partial_t q^a(x) &=&  \int d^3x' \; \left[ N {\d \over \d p_a(x)}\H
                            + N_i {\d \over \d p_a(x)}\H^i \right]
\non \\
  \partial_t p_a(x) &=& -  \int d^3x'\; \left[ N{\d \over \d q^a(x)}\H
                            + N_i {\d \over \d q^a(x)}\H^i \right]
\non \\
      \H &=& 0
\non \\
      \H^i &=& 0
\label{H1}
\eea
As pointed out by Moncrief and Teitelboim \cite{MT},
the supermomentum constraints $\H^i=0$ need not be imposed
independently.  These constraints are implied
by the requirement that the Hamiltonian constraints $\H=0$ are preserved
by the time evolution, which demands the vanishing of the Poisson
brackets $\{\H(x),\H(y)\}$.  Since these Poisson brackets turn out to
be linear in $\H^i$, the momentum constraint follows.

   To help fix the notation, we note that for pure gravity, where
the conjugate variables are the three metric $g_{ij}$ and corresponding
momenta $p^{ij}$, the various expressions in the ADM action are as follows:
\bea
        \{a=1-6\} &\leftrightarrow& \{(i,j),~i \le j \}
\non \\
             q^a(x) &\leftrightarrow& g_{ij}(x)
\non \\
             p_a(x) &\leftrightarrow&
\left\{ \begin{array}{rr}
             p^{ij}(x)~~~~~(i=j) \\
           2 p^{ij}(x)~~~~~(i<j) \\
        \end{array} \right.
\non \\
             G_{ab}(x) &\leftrightarrow&  G^{ijnm}(x)
\non \\
          \rg U &=& - {1 \over \k^2}\rg {~}^3R
\non \\
            \H^i &=& -2p^{ik}_{~~;k}
\eea
where $G_{ijkl}$ is the DeWitt superspace metric
\beq
   G_{ijkl} = {1 \over 2\rg}(g_{ik}g_{jl}+g_{il}g_{jk}-g_{ij}g_{kl})
\eeq
In this case, the Hamilton equations plus constraints \rf{H1}
are equivalent to the vacuum Einstein equations.

   The lapse and shift functions $N$ and $N_i$ foliate spacetime
into space + time, and set the coordinates on each constant-time
hypersurface.  If one sets $N_i=0$, it is still possible to
choose arbitrary foliations using the lapse function, although the
coordinates on each time-slice are then fixed.  It is also possible,
without affecting the freedom to choose arbitrary foliations, to
limit the lapse functions $N$ to a subset $N=\tN$ satisfying
\beq
       \int d^3x \; \tN \rg U(q) = \s
\label{tN}
\eeq
where $\s$ is an arbitrary parameter with dimensions of mass.
Equivalently,
\beq
        \tN = {\s \N \over \int d^3x \N \rg U(q)}
\label{tN1}
\eeq
where $\N$ is unconstrained.  The global constraint \rf{tN}, like the choice
$N_i=0$, does not limit the choice of constant-time hypersurfaces; it
only affects the value of the label $t$ assigned to each hypersurface.
Making the gauge choices
\beq
        N_i=0 ~~~~~ \mbox{and} ~~~~~ N=\tN
\label{gauge}
\eeq
the 1st-order equations of motion and constraints become
\bea
     \pa_t q^a(x) &=& 2\tN \k^2 G^{ab}p_b(x)
\non \\
     \pa_t p_a(x) &=& - \int d^3x' \tN {\d \over \d q^a(x)} \H
\non \\
      \H &=& \k^2 G^{ab} p_a p_b + \rg U = 0
\label{H2}
\eea
The supermomentum constraints have been dropped, since they are
implied by the other equations.  We go to the 2nd-order form
by solving the first of these equations for $p_a$
\beq
        p_a = {1 \over 2 \tN \k^2} G_{ab} \pa_t q^b
\eeq
and inserting into the second two equations of \rf{H2}, to get
\beq
  {\pa \over \pa t}\left[ {1\over 2\tN \k^2} G_{ab} \pa_t q^b \right]
 - {1 \over 4\tN \k^2}{\pa G_{cd} \over \pa q^a}\pa_t q^c \pa_t q^d
+ \int d^3x' \; \tN {\d \over \d q^a(x)}(\rg U) = 0
\label{motion}
\eeq
and
\beq
  {1 \over 4\tN^2 \k^2} G_{ab}\pa_t q^a \pa_t q^b + \rg U = 0
\label{constraint}
\eeq
as the classical field equations with lapse/shift conditions \rf{gauge}.

   Let us now introduce some notation.  Define a mixed discrete/continuous
index $(\a,x)$ as a "coordinate index" in superspace
\bea
     q^{(\a x)} \equiv q^{\a}(x) = \left\{ \begin{array}{cl}
                     \N(x)   & \a=0 \\
                     q^a(x)  & \a=a \ne 0 \end{array} \right.
\eea
Apart from notation, we are enlarging the definition of superspace to include
the field $\N(x)$, which appears in eq. \rf{tN1}.  The summation convention
for tensor indices is then
\beq
       V_{..(\a x)..} W^{..(\a x)..} \equiv \sum_\a \int d^3x \;
             V_{..(\a x)..} W^{..(\a x)..}
\eeq
We are now ready to state the main results:

\begin{description}

\item{I.~} The equation of motion \rf{motion} and Hamiltonian constraint
\rf{constraint} are the equations of a
geodesic in superspace, with the time label $t$ an affine parameter in
superspace, proportional to the proper-time along the geodesic;

\item{II.} Both the equation of motion \rf{motion} and the constraint
\rf{constraint} are obtained by extremizing the proper-time of the
path in superspace, such that the affine parameter $t$, given by
\beq
   t = {1\over \s}\int d\t \; \sqrt{-\G_{(\a x)(\b y)}
           {dq^{(\a x)} \over d \t}  {dq^{(\b y)} \over d \t} }
\label{proper}
\eeq
is stationary with respect to variations of $q^{(\a x)}(\t)$, where the
(degenerate) metric of superspace is proportional to
\beq
  \G_{(\a x)(\b y)} = \left\{ \begin{array} {cl}
          \left[\int d^3x' \; \N \rg U \right]
          {1 \over 4\N(x) \k^2} G_{ab}(x) \d^3(x-y) & \a = a,~~\b = b \\
                   0 & \a=0 ~~\mbox{and/or}~~\b=0 \end{array} \right.
\label{metric}
\eeq
\end{description}

   The equation of motion \rf{motion} is obtained from the
stationarity condition
\beq
       {\d t \over \d q^{(ax)}(\t)} = 0
\eeq
with discrete index $a \ne 0$.
Taking the indicated functional derivative of \rf{proper}, we find
\bea
  0 &=& {d \over d \t} \left\{ {\V \over \K} {1 \over 2 \N \k^2} G_{ab}
             {dq^b \over d\t} \right\}
\non \\
    &-& {\V \over \K} {1 \over 4 \N \k^2} {\pa G_{cd} \over \pa q^a}
               {dq^c \over d\t}{dq^d \over d\t}
\non \\
    &+&  {\K \over \V} \int dx''' \; \N {\d \over \d q^a} (\rg U)
\eea
Defining $\tN$ according to eq. \rf{tN1}, this becomes
\bea
0 &=&   {d \over d\t} \left\{ {1 \over {1\over \s}\V \K} {1 \over 2\tN \k^2}
     G_{ab} {dq^b \over d\t} \right\}
\non \\
  &-& {1 \over {1\over \s}\V \K} {1\over 4\tN \k^2}{\pa G_{cd} \over \pa q^a}
      {dq^c \over d\t}{dq^d \over d\t}
\non \\
  &+&  {1\over \s}\V \K \int dx''' \; \tN {\d \over \d q^a} (\rg U)
\non \\
\eea
and then using
\bea
dt &=& {1\over \s} \sqrt{-\G_{(\a x)(\b y)}{dq^{(\a x)} \over d\t}
         {dq^{(\b y)} \over d\t} } d\t
\non \\
   &=& {1\over \s} \V \K d\t
\label{dt_dtau}
\eea
we obtain
\beq
   {d \over dt} \left\{ {1\over 2\tN \k^2} G_{ab} {d q^b \over dt} \right\}
       - {1\over 4\tN \k^2} {\pa G_{cd} \over \pa q^a}
      {dq^c \over dt}{dq^d \over dt}
       + \int dx' \; \tN {\d \over \d q^a} (\rg U) = 0
\eeq
which is identical to the classical equation of motion \rf{motion}.

   The Hamiltonian constraint is obtained from the remaining stationarity
condition
\beq
    {\d t \over \d q^{(0x)}(\t) } = {\d t \over \d \N(x,\t)}= 0
\eeq
Again taking the indicated functional derivative gives
\bea
\lefteqn{   {\V \over \K}{1 \over 4\N^2 \k^2} G_{ab}\pa_\t q^a \pa_\t q^b }
\non \\
    & &   + {\K \over \V} \rg U = 0
\eea
and using \rf{tN1} and \rf{dt_dtau} we find
\beq
  {1\over 4\tN^2 \k^2}G_{ab}{dq^a \over dt}{dq^b \over dt}
       + \rg U = 0
\eeq
which is simply the Hamiltonian constraint \rf{constraint}.  Consistency
of the Hamiltonian constraint with the equations of motion \rf{motion}
then implies the supermomentum constraints.  In this way, the full
set of classical field equations and constraints are obtained from
the requirement
\beq
        {\d t \over \d q^{(\a x)}(\t)} = 0
\eeq
verifying the results (I) and (II) stated above.  Note that equations
\rf{motion} and \rf{constraint} are covariant with respect to a change
in the mass parameter $\s$; the increment of proper
time on the 4-manifold, namely $\tN dt$, is $\s$-independent.
The constant $\s$ is completely arbitrary, and has been introduced
only to give the evolution parameter $t$ the dimensions of time.
{}From the point of view of the
ADM equations of motion, a change of $\s$ is just a relabeling
of the time variable $t$; from the point of view of the geodesic condition,
it is simply a rescaling of the affine parameter.

  If, instead of using the explicit form \rf{metric} of the
supermetric $\G_{(\a x)(\b y)}$, the supermetric is left arbitrary,
then variation of $t$ by $q^\a(x)$ leads, by standard manipulations,
to the equation
\beq
     \G_{(\a x)(\b y)} {d^2 q^{(\b y)} \over d t^2} + \oh \left(
     {\d \G_{(\a x)(\b y)} \over \d q^{(\gamma z)} } +
     {\d \G_{(\a x)(\gamma z)} \over \d q^{(\b y)} } -
     {\d \G_{(\b y)(\gamma z)} \over \d q^{(\a x)} } \right)
     {dq^{(\b y)} \over d t}{dq^{(\gamma z)} \over d t}
     = 0
\label{geo}
\eeq
Inserting the supermetric \rf{metric}, it is not difficult to verify
explicitly that the $\a=0$ component of this equation is the
Hamiltonian constraint \rf{constraint}, while the $\a =a \ne 0$
components are just the field equations \rf{motion}.
If the metric were invertible, one could multiply this expression
by the reciprocal metric, and obtain the usual form of the geodesic
equation
\beq
      {d^2 q^{(\a x)} \over dt^2} + \Gamma^{(\a x)}_{(\b y)(\gamma z)}
            {d q^{(\b y)}\over dt}{d q^{(\gamma z)}\over dt} = 0
\label{geodesic}
\eeq
where $\Gamma^{(\a x)}_{(\b y)(\gamma z)}$ is the connection
corresponding to the supermetric $\G_{(ax)(by)}$.
This supermetric is not invertible, however, and there is no term
involving a second derivative $d^2 \N /dt^2$.  As a result, there is
not one, but rather a continuously infinite set of trajectories
$\{ q^a(x,t),\N(x,t) \}$ which extremize the proper time in superspace,
proportional to the affine parameter $t$, between
given initial $q_{in}^{(\a x)}$ and final $q_f^{(\a x)}$
configurations.  This is as it should be.  Each geodesic (a solution of
the field equations), satisfying given initial and final boundary conditions,
represents a different foliation, corresponding to a particular choice
of lapse, of a certain 4-manifold. The set of all such geodesics, corresponding
to all possible foliations of the same 4-manifold, forms
an equivalence class.\footnote{There may be more than one equivalence
class, since there many be more than one 4-manifold solving the equations
of motion, bounded by $q_{in}$ and $q_f$.} Thus the non-invertibility of
the supermetric, which leads to an infinite degeneracy in solutions for a
geodesic,
is just a consequence of the (ordinary) time diffeomorphism invariance in
four-dimensional space, which allows for an infinite number of possible
foliations.

  If a set of stationary paths between two points in superspace are just
different representations of the same 4-manifold (+ non-gravitational
fields), then one would expect
that the "proper time" interval in superspace along each path would be the
same.
In fact, the proper time in superspace, along a geodesic
joining two configurations $\{q_{in}^a(x)\}$ and $\{q_f^a(x)\}$ is proportional
to the diffeomorphism invariant action
\beq
     S = \int d^4x \; \left[ -{1\over \k^2}\rg R
             + {\cal L}_{non-grav}(g_{\m\n},\phi^A) \right]~~+~~
            \mbox{boundary terms}
\label{action}
\eeq
evaluated on a solution of the equations of motion bounded by the
given initial and final configurations. To derive
the proportionality of $t$ and $S$,
begin with the definition of the affine parameter $t$ in superspace
\bea
      -1 &=& {1\over \s^2}\G_{(ax)(by)} {dq^{(ax)} \over dt}{dq^{(by)} \over
dt}
\non \\
         &=& {1 \over \s} \int d^3x \; {1 \over 4 \tN \k^2} G_{ab}
                 {dq^a \over dt}{dq^b \over dt}
\label{minus}
\eea
Solving the Hamiltonian constraint \rf{constraint} for $\tN$ gives
\beq
     \tN = \sqrt{ - {1 \over 4\k^2 \rg U} G_{ab}
                   {dq^a \over dt}{dq^b \over dt} }
\eeq
and substituting this expression into \rf{minus} we have
\beq
     1 = {1\over \s} \int d^3x \sqrt{ - {1 \over 4\k^2} \rg U G_{ab}
                 {dq^a \over dt}{dq^b \over dt} }
\eeq
or
\beq
     dt = {1\over \s} \int d^3x \; \sqrt{ - {1 \over 4\k^2}
                         \rg U G_{ab} dq^a dq^b  }
\eeq
Then
\beq
      t = {1\over \s} \int d^4x \; \sqrt{ - {1 \over 4\k^2} \rg U G_{ab}
             {dq^a \over dx^0} {dq^b \over dx^0}  }
\label{BSW}
\eeq
The integral in eq. \rf{BSW} is the Baierlein-Sharp-Wheeler form of the
action \rf{action}, in "shift gauge" $N_i=0$; it is obtained
from the ADM action by solving for the momenta and the lapse function
in terms of the velocities \cite{BSW}.
The value of the BSW action, evaluated along a stationary path,
is equal to the diffeomorphism-invariant action \rf{action}
evaluated along the same path.  Because of diffeomorphism invariance,
any geodesic in an equivalence class, subject to given initial
and final boundary conditions, will have the same action $S$.  Therefore,
since
\beq
        t = {1 \over \s} S
\eeq
evaluated along the geodesic, all geodesics between given end-points
in superspace, which differ (in 4-space) only by a foliation, have the
same interval of proper time in superspace.

\section{Jacobi's Principle}

   The stationarity of $t$ in \rf{proper} is closely related
to Jacobi's principle in classical non-relativistic mechanics.
Consider a particle of energy $E$, whose motion in a D-dimensional
space is governed by the Hamiltonian
\beq
     H = {1\over 2m} K^{ab}(x)p_a p_b + V(x)
\eeq
According to Jacobi's principle, the path traced out in $x$-space by
the particle trajectory $x(\t)$ extremizes the quantity
\beq
       s =  \int d\t \sqrt{ \G_{ab} {dx^a \over d\t}{dx^b \over d\t} }
\label{s}
\eeq
where
\beq
       \G_{ab} = (E-V) m K_{ab}
\label{Jmetric}
\eeq
and where $K_{ab}$ is inverse to $K^{ab}$.  The similarity of these
expressions to \rf{proper} and \rf{metric} is obvious.
Despite these similarities, one would not say that a
non-relativistic particle moving in an arbitrary potential is in free fall.
For one thing, the Euler-Lagrange equations for a non-relativistic particle
involve the mass parameter $m$, while the geodesic equation derived from
Jacobi's principle does not.  Jacobi's principle only concerns
the parametrized orbit $x^a(\t)$, while the Euler-Lagrange equations deal
with the trajectory $x^a(T)$ in terms of the Newtonian time $T$.  To obtain
the Euler-Lagrange equations, it is necessary to introduce an additional
condition
\beq
      \oh m K_{ab}{dx^a \over dT}{dx^b \over dT} + V = E
\label{Newton}
\eeq
which defines the Newtonian time $T$.  The mass parameter enters the equations
of motion at this point.  Since the geodesic equation by itself is {\it not}
equivalent to the Euler-Lagrange equations, the motion of a non-relativistic
particle in an arbitrary potential is not equivalent to free fall.

  On the other hand, consider the relativistic action
\beq
       S = -m \int d\t \sqrt{ -g_{\m \n}(x) {dx^\m \over d\t}{dx^\n \over d\t}
}
\label{sr}
\eeq
In this case, the parametrized trajectory $x^\m(\t)$ contains all there is
to know about the particle's motion; the Euler-Lagrange equations and the
geodesic equation are equivalent, and do not involve the particle mass $m$.
The action \rf{sr} therefore describes a particle in free fall.

   Free fall, for a relativistic particle, can be reformulated as the
motion of a particle in a certain kind of potential.  To see this,
consider an arbitrary factoring of $g_{\m \n}$ into two parts
\beq
       g_{\m \n}(x) = \phi(x) G_{\m \n}(x)
\eeq
Then the first-order form of the action \rf{sr} is given by
\bea
       S &=& \int d\t \left[ p_\m \pa_\t x^\m - N\H \right]
\non \\
      \H &=& {1\over 2m} G^{\m \n} p_\m p_\n + \oh m \phi(x)
\label{fo}
\eea
This is verified by first solving for the momenta in terms of the
velocities using $\pa_\t x^\m = N\pa \H / \pa p_\m$, then solving
for the lapse function $N$
from the Hamiltonian constraint $\H=0$, and finally substituting
the results into the first-order action in \rf{fo}. The square-root action
\rf{sr} follows. Note, however, that the Hamiltonian constraint can be
written as
\beq
        G^{\m \n} p_\m p_\n + m^2 \phi(x) = 0
\eeq
This equation can be interpreted as referring to a particle moving in a
spacetime of background metric $G_{\m \n}$, with a "position-dependent mass"
$m\sqrt{\phi(x)}$.  Therefore, the motion of a particle in a manifold of
metric $G_{\m \n}$, with a position-dependent mass, is simply a different
way of formulating the free fall of particle in a manifold of metric
$g_{\m \n}$, with a constant mass.  Note again the similarity to the
Jacobi form \rf{s} and \rf{Jmetric}.  In this case, $\G_{ab},~K_{ab},~
mV(x)$ in \rf{Jmetric} correspond to $g_{\m \n}, ~G_{\m \n}, ~\phi(x)$,
respectively, with $E=0$.

  Now in gravitation theory, the metric of superspace is usually taken to be
the ultralocal DeWitt metric
\beq
     G_{(ax)(by)} = G_{ab}[g_{ij}(x)] \d^3(x-y)
\eeq
and the term $\rg U$ in the Hamiltonian is viewed as the
potential.  In close analogy to the position-dependent mass term of a
relativistic particle, the potential term in the ADM Hamiltonian can
be absorbed into a redefinition of the metric (eq. \rf{metric}), and we
arrive at the action principle \rf{proper}.  Taking
$\G_{(\a x)(\b y)}$, rather than $G_{(ax)(by)}$, as the supermetric,
the dynamical field equations decribe the free fall of a point particle
through superspace.  As in the case of the relativistic particle,
the geodesic equation is equivalent to the classical equations of motion,
and a given geodesic $q^\a(x,t)$ through superspace provides a complete
description of the dynamics.  No additional information, analogous
to the definition of Newtonian time in \rf{Newton}, is required, and no
constants (analogous to mass) not appearing in the geodesic equation
are needed.

    It should be noted that there has been
some previous work on the topic of Jacobi's principle
in general relativity, by Brown and York \cite{BY}.  In their approach
the cosmological constant is interpreted as being
analogous to an energy parameter.
This entails a modification of classical relativity (the unimodular theory
\cite{unimod}) in which the
cosmological constant is taken to be a dynamical degree of freedom,
much like the energy $E$ of a non-relativistic theory, which is
conjugate to another variable interpreted as a time-evolution
parameter.  The motivation for introducing such a parameter was to address
the problem of time in quantum gravity.  As in the non-relativistic case,
the equations of motion in this particular time parameter are not geodesic
equations.  The reader is referred to ref. \cite{BY} for further details
concerning this approach.

\section{Quantization}

   We next consider the first-order formulation of free fall in superspace;
the object is to find the quantity whose Poisson brackets evolve the system
in the affine parameter $t$.

   In ordinary 4-space, the geodesic equation is obtained by variation of
\beq
       s = \int d\t \; \sqrt{-g_{\m \n} \pa_\t x^\m \pa_\t x^\n}
\eeq
with respect to $x^\m$, and then using
\beq
       ds =  \sqrt{-g_{\m \n} \pa_\t x^\m \pa_\t x^\n} d\t
\label{ds}
\eeq
to replace the arbitrary parameter $\t$ with the proper-time $s$ in the
resulting equations of motion.   As is well known \cite{MTW}, the geodesic
equation is also obtained by varying the action
\beq
       S = m_0 \int dt \; \oh g_{\m \n} {dx^\m \over dt} {dx^\n \over dt}
\label{2nd}
\eeq
with respect to $x^\m(t)$.  This leads directly to a geodesic equation
\beq
      {d^2 x^\m \over dt^2} + \Gamma^\m_{\a \b}{dx^\a \over dt}
                                                {dx^\b \over dt} = 0
\eeq
where the affine parameter $t$ is proportional to the interval in proper
time $s$.  Going to a first-order formulation of the action \rf{2nd}
\beq
       S = \int dt \; \left[ p_\m \pa_t x^\m - {1\over 2m_0}
                              g^{\m \n} p_\m p_\n \right]
\eeq
we see that the quantity
\bea
       \E &\equiv& - { p^2 \over m_0^2 }
\non \\
          &=&  - g_{\m \n} {dx^\m \over dt}{dx^\n \over dt}
\eea
is a constant of motion, and therefore, using \rf{ds},
\beq
          ds = \sqrt{\E} dt
\eeq
This is the relationship between the evolution parameter $t$,
and the proper time $s$.  Note that the action \rf{2nd}
has no constraint on the mass $m^2=-p^2=\E m_0^2$ of the particle, which is
treated simply as a constant of motion.  This mass, of course, cannot
be determined from the particle trajectory (a geodesic) in configuration
space.

In the same way, the equations of motion \rf{motion} and constraint
\rf{constraint} in superspace are obtained by variation of the action
\bea
     S &=& {1 \over \s_0} \int dt \; \G_{(\a x)(\b y)}
             {dq^{(\a x)} \over dt} {dq^{(\b y)} \over dt}
\non \\
       &=& {1 \over \s_0} \int dt \; \G_{(a x)(b y)}
             {dq^{(a x)} \over dt} {dq^{(b y)} \over dt}
\eea
Going over to the first-order formulation in the usual way, we find
\beq
     S = \int dt \; \left( p_{(ax)}{dq^{(ax)} \over dt} - \A \right)
\eeq
where
\beq
        \A = \oq \s_0 \G^{(ax)(by)}p_{(ax)}p_{(by)}
\eeq
and
\beq
        \G^{(ax)(by)} = {1 \over \int d^3x' \; \N \rg U}
                           4\N \k^2 G^{ab} \d^3(x-y)
\eeq
The quantity $\A$ is a constant of motion, denoted $\A = -\E \s_0$, of
the corresponding dynamical equations
\beq
    {dq^{(ax)} \over dt} = {\d \A \over \d p_{(ax)}}~~,~~~~~~~
    {dp_{(ax)} \over dt} =-{\d \A \over \d q^{(ax)}}~~,~~~~~~~
    0 = {\d \A \over \d \N(x)}
\label{AE-eq}
\eeq
It is straightforward to verify that eq. \rf{AE-eq}
reproduces the equations of motion \rf{motion} and constraint
\rf{constraint}, upon solving for the momenta in terms of velocities,
and identifying $\s = \ER \s_0$ in $\tN$.

   In contrast to the usual Hamiltonian of general relativity, the
quantity $\A[q,p,\N]$ is {\it not} constrained to be zero, although
the constraint equation \rf{constraint} is obtained from variation
with respect to $\N$.  The value $\A = -\s_0 \E$ is a constant of motion,
which, however, cannot be determined from the trajectory followed in
superspace.  For $\A = -\s_0 \E$, the constraint $\d \A / \d \N = 0$
is equivalent to
\beq
      \H^\E_x = {\k^2 \over \ER} G^{ab} p_a p_b + \ER \rg U = 0
\label{H-E}
\eeq
This looks like the usual Hamiltonian constraint, apart from
the presence of a free parameter $\E$.  In fact,
the parameter $\E$ is irrelevant to the configuration-space
equations of motion.  This is seen by starting from the Hamiltonian
equations of motion based on $\H^\E_x$,
\bea
      {dq^{(ax)} \over d\t} &=& \int d^3x' \; N(x')
                       {\d \H^\E_{x'} \over \d p_{(ax)}}
\non \\
      {dp_{(ax)} \over d\t} &=& - \int d^3x' \; N(x')
                           {\d \H^\E_{x'} \over \d q^{(ax)}}
\non \\
            \H^\E_x &=& 0
\eea
and solving for the momenta in terms of velocities.  One then
finds that the parameter
$\E$ drops out of the resulting second-order equations of motion
and constraint.  This means that the constant $\E$, like the mass
of a particle in free fall, or the tension of the Nambu string,
or the value of Newton's constant in vacuum gravity,\footnote{In vacuum
gravity, where $\rg U = -\rg R/\k^2$, the undetermined constant $\E$ can
be absorbed, by a scaling $\k^2_{new}=\k^2/\ER$, into an
(undetermined) Newton's constant.}
doesn't appear in the Euler-Lagrange
equations of the theory, and hence cannot be determined from the
classical trajectories in configuration space.  In this sense,
$\E$ is an undetermined constant. The existence and implications
of such constants in Hamiltonian constraints has been discussed
in ref. \cite{Us1}.

   The dynamical equations \rf{AE-eq} can be rewritten in Poisson
bracket form
\beq
{d \over dt}Q[q,p] = \{ Q,\A \} ~,~~~~~ {\d \A \over \d \N(x)} = 0
\label{PB}
\eeq
Since $\A$ is a constant of motion whose value at the classical level
is unconstrained, and since the equations  \rf{AE-eq} are equivalent
to the field equations of general relativity,\footnote{The field
equations are in $N_i=0$ gauge, but with no restriction on foliation.}
it seems reasonable to base the canonical quantization of gravity on the
Schrodinger equation corresponding to the Poisson bracket of
eq. \rf{PB}, i.e.
\beq
  i\hbar {d \Psi \over dt} = \A \Psi ~~~~~~~~~\mbox{(any $\N$)}
\label{S-eq}
\eeq
thereby avoiding the notorious "problem of time"
in quantum gravity \cite{Kuchar}.
In fact, this proposal has been made and developed by Carlini and the
author in a series of papers \cite{Us2}-\cite{Me}, following a different
line of reasoning.     One of the advantages of this proposal, discussed
in more detail in the cited references, is that it allows for the existence of
physical states which are eigenstates of non-stationary observables.
Because a physical state can only depend on the configuration-space variables
$\{ q^a(x) \}$, and not on $\N(x)$, such states must
satisfy \rf{S-eq} for any choice of $\N$.  Expanding a time-dependent
solution in terms of stationary states
\beq
\Psi[q,t] = \sum_{\E,\a} c_{\E,\a} \Phi_{\E,\a}[q]
               e^{i \s_0 \E t /\hbar}
\eeq
and inserting into \rf{S-eq}, the condition of $\N$-independence
implies
\beq
      \left[ -{\hbar^2 \over \E} G^{ab}{\d^2 \over \d q^a \d q^b}
                + \rg U \right]  \Phi_{\E,\a}[q] = 0
\eeq
which is a Wheeler-DeWitt equation (with the usual ordering ambiguity)
parametrized by $\E$.  This is the operator version of the
constraint \rf{H-E} corresponding to \newline
$\d \A / \d \N = 0$;
the label $\a$ distinguishes among a
linearly independent set of solutions of this equation.
The physical Hilbert space is thereby spanned by the solutions of a
one-parameter ($\E$) {\it family} of Wheeler-DeWitt equations.
A solution of any given Wheeler-DeWitt equation, with fixed $\E$, is
a stationary state; it cannot be an eigenstate of non-stationary
observables, such as 3-geometry or extrinsic curvature.  However, a
superposition of states $\Phi_{\E,\a}$, with different $\E$, is a
non-stationary state. Such states can indeed be eigenstates of
non-stationary observables, and therefore have the possibility of,
e.g., describing the outcome of a measurement process.  All of this
has been discussed in some detail in ref. \cite{Us2}-\cite{Me}.
But what we now see, from the work of the previous section,
is that {\it the time parameter for quantum gravity in
eq. \rf{S-eq} is, at the classical level, proportional to
the proper-time of the trajectory of the Universe in superspace}.

\section{Beyond Free Fall}

   The geodesic equation of motion in general relativity is the
statement that the non-gravitational force on a particle is zero.
The equivalent statement, in superspace, is that all dynamics is free fall;
there no other forces in superspace that act on the Universe, viewed as
a point particle.  We have no motivation, from phenomenology,
to go beyond this statement.  Still, it is intriguing to consider
what might be the form of the equations of motion if there {\it would} be
some non-(super)gravitational forces in superspace, presumably so weak
as to have gone undetected.  In other words, what is the analog
of $F=ma$ in superspace, and are there any consistency conditions
that must be satisfied by such "superforces"?

   The motion of a point particle in ordinary spacetime is governed by
the equation
\beq
      g_{\m\n} {d^2 x^\n \over d\t^2} + \oh \left(
             {\pa g_{\m \a} \over \pa x^\b}
           + {\pa g_{\m \b} \over \pa x^\a}
           - {\pa g_{\a \b} \over \pa x^\m} \right)
      {dx^\a \over d\t}{dx^\b \over d\t}
             = {1\over m} F_\m
\eeq
The direct generalization to superspace is an equation of the form
\bea
     \G_{(\a x)(\b y)} {d^2 q^{(\b y)} \over d t^2} &+& \oh \left(
     {\d \G_{(\a x)(\b y)} \over \d q^{(\gamma z)} } +
     {\d \G_{(\a x)(\gamma z)} \over \d q^{(\b y)} } -
     {\d \G_{(\b y)(\gamma z)} \over \d q^{(\a x)} } \right)
     {dq^{(\b y)} \over d t}{dq^{(\gamma z)} \over d t}
\non \\
     &=& \r F_{(\a x)}
\label{F=ma2}
\eea
where $\r$ is a constant, and $F_{(\a x)}$ is the "superforce."
As before, the $\a = 0$ components of this equation are the equations
of constraint, while the $\a \ne 0$ components are the equations
of motion.  The requirement that the equations of constraint are
preserved by the motion leads to certain conditions on the form
of the superforce.

   In ordinary spacetime, the electromagnetic force acting on a charged
particle preserves the mass-shell condition; i.e. the
Hamiltonian constraint
\beq
       H = {1\over 2m} p^2 + \oh m = 0
\eeq
is unchanged.  Let us assume that
this is also true in superspace, which would insure that the number of
independent dynamical degrees of freedom is unaffected by the force term.
This requires $F_{(0x)}=0$.  Then eq. \rf{F=ma2} for $\a = a \ne 0$ becomes
\beq
  {d \over dt}\left[ {1\over 2\tN \k^2} G_{ab} { dq^b \over dt}\right]
 - {1 \over 4\tN \k^2}{\pa G_{bc} \over \pa q^a} {dq^b \over dt}
{dq^c \over dt} + \int d^3x' \; \tN {\d \over \d q^a(x)}(\rg U)
        = {2\r \over \s} F_{(ax)}
\label{A2}
\eeq
while the constraint, obtained from \rf{F=ma2} with $\a = 0$,
\beq
  {1 \over 4\tN^2 \k^2} G_{ab}\pa_t q^a \pa_t q^b + \rg U = 0
\label{A1}
\eeq
is unchanged.

   Again define
\beq
      \H_x = \k^2 G^{ab} p_a p_b + \rg U
\eeq
Then the equation of motions \rf{A2} and constraint \rf{A1} are
equivalent to
\bea
      {dq^{(ax)} \over dt} &=& \int d^3x' \; \tN(x') {\d \H_{x'} \over \d
p_{ax}}
\non \\
      {dp_{(ax)} \over dt} &=& - \int d^3x' \; \tN(x')
{\d \H_{x'} \over \d q^{ax}}  + {2\r \over \s} F_{(ax)}
\non \\
            \H_x &=& 0
\label{B}
\eea
Consistency of the Hamiltonian constraint $\H_x=0$ with the other
equations of motion in \rf{B} then requires
\bea
      0 &=& {d\H_x \over dt}
\non \\
        &=& {\d \H_x \over \d q^{(ax')}} {dq^{(ax')} \over dt}
         +  {\d \H_x \over \d p_{(ax')}} {dp_{(ax')} \over dt}
\non \\
        &=& \int d^3x' \; \tN(x') \{ \H_x,\H_{x'} \}
              ~+~ {4\r \k^2 \over \s} G^{ab} p_a(x) F_b(x)
\eea
Since the Poisson Bracket $ \{ \H_x,\H_{x'} \} $ is linear in the supermomentum
density $\H^i_x$, consistency is obtained by imposing the usual
supermomentum constraint
\beq
          \H^i_x = 0
\eeq
on the canonical momenta, as well as the ``orthogonality condition'' on
the force
\beq
       p_a(x) F^a(x) \equiv G^{ab} p_a(x) F_b(x)  = 0
\eeq
at each point $x$.
But then, we also need to ensure that the supermomentum constraint is
maintained by the equation of motion.  This requires
\bea
      0 &=& {d \H^i_x \over dt}
\non \\
        &=& \int d^3x' \; \tN(x') \{ \H^i_x,\H_{x'} \}
  ~+~ {2\r \over \s} {\d \H^i_x \over \d p_{(ax')}} F_{(ax')}
\eea
The vanishing of the Poisson bracket $\{ \H^i_x,\H_{x'} \} $ is guaranteed
by the usual Dirac algebra, and $\H_x=0$, so the above condition reduces
to a supermomentum constraint on the force term
\beq
        \H^i_x \left[ p_a(x) \ra F_a(x) \right] = 0
\eeq
where the notation means that the momenta in the supermomentum
constraint are replaced by the corresponding components of the
superforce. In short, a consistent force term in the superspace equations
of motion which preserves the form of the constraints must:
(1) be velocity-dependent, in the sense that the force
is orthogonal to the canonical momenta at every point in 3-space; and
(2) obey a supermomentum constraint. Altogether:
\bea
              p_a(x) F^a(x) &=& 0
\non \\
 \H^i_x \left[ F_a(x) \right] &=& 0
\eea
The first of these conditions is reminiscent of the Lorentz force of
electromagnetism,
\beq
         F^\m = e g^{\m \n} (\pa_\n A_\l - \pa_\l A_\n) {dx^\l \over ds}
\eeq
which is, in fact, orthogonal to the particle 4-momentum, i.e. $p_\m F^\m = 0$,
in ordinary spacetime.  This orthogonality is due to the
antisymmetry of the electromagnetic field tensor, and it guarantees
that the electromagnetic force leaves the 4-momentum of a charged particle
on its mass shell: $H=0 ~ \Rightarrow ~ p^2 + m^2 = 0$.  The second condition,
a super-momentum constraint on the force, has no obvious analog in
particle dynamics.

\section{Conclusions}

   The fact that the Einstein (+ other integer-spin) field equations can
be reformulated as a single geodesic equation may be of interest
aesthetically. More importantly, the formalism also suggests a natural
evolution operator (and evolution parameter) for the corresponding quantum
theory, which has obvious application to the problem of time.  It would be
interesting to see if this geodesic reformulation can also be extended
to include spinor fields.

   Beyond this, it is tempting to speculate that superspace should be
regarded, rather literally, as a true arena of dynamics. The Universe is a
point particle in this space,
and it moves, under the influence of a (super)gravitational field, along
a geodesic.  If one particle can fall freely in superspace, why not
others?  Interactions among such particles would result in
deviations from geodesic motion, as discussed in the last section.
This suggests that the geodesic equation in superspace, and its possible
extensions, might be a natural starting point for constructing classical
theories of multi-universe dynamics. The many-universe concept
is not new in quantum gravity; in particular, it has been argued that
wormhole processes are best described in the framework of third
quantization \cite{GS}.  A classical theory of multi-universe dynamics, if
it could be constructed consistently, might well be the
"particle limit" (in the sense of ref. \cite{Marty}) of a corresponding
third-quantized field theory, associated with topology-changing processes.

\vspace{33pt}

\noindent {\Large \bf Acknowledgements}

\bigskip

I am grateful for the hospitality of the theory group at the Lawrence
Berkeley Laboratory, where much of this work was carried out; I would
also like to thank Alberto Carlini for helpful discussions.
This work is supported in part by  the U.S. Department of Energy, under
Grant No. DE-FG03-92ER40711; support has also been provided by the
Danish Natural Science Research Council.

\end{document}